\documentclass{pnaoj08}

\begin{document}
\SetRunningHead{Miyoshi M.}{Re-analysis of the First Fringe with 2-Beam in the VERA System from Archive Data}

\Received{2008/10/9}
\Accepted{2008/12/16}

\title{Re-analysis of the First Fringe 
with 2-Beam \\
in the VERA System from Archive Data}

\author{Makoto \textsc{Miyoshi},\altaffilmark{1}}
 \altaffiltext{1}{National Astronomical Observatory, Mitaka, Tokyo, 181-8588, Japan.}\email{makoto.miyoshi@nao.ac.jp}



\maketitle

\begin{abstract}
We report results from re-analysis of the visibility data of the first 2-beam observations with VERA (VLBI Exploration of Radio Astrometry), previously reported by Honma et al., 2003 (hereafter A2003). Independently we checked the archival data and found the features noted in A2003 were not from the effect of phase referencing by simultaneous differential VLBI but mainly from a removal of large phase change by subtracting an arbitrary fitted curve to the phase variations.

The differential phase of the observed H$_2$O masers between W49 North (W49N) and OH~43.8-0.1 did not show a sinusoidal variation with a period of one sidereal day due to a positional offset from the real celestial positions. 
 We therefore could not reproduce the results in A2003 by a normal positional correction estimated from all time data, but could reproduce almost the same phases only for the first hour by adjusting parameters. Using the parameters, we could not suppress the large amount of phase variations for the successive time data that A2003 did not show in their paper.
 It is appropriate to regard the analysis in A2003 as not being proper for showing  the performance of the instrument for phase referencing, which should be demonstrated by other experiments observing several pairs of continuum sources.


\end{abstract}

\section{Introduction}
 The Japanese VLBI project VERA has a long history originating in the former ILOM (International Latitude Observatory of Mizusawa). With passing time the target of the project changed from frequent geodetic VLBI monitoring of earth rotation to researches on Galactic structure and dynamics using maser astrometry (Hara 1986, Fujishita \& Hara 1988, Hara 1988, Sasao \& Morimoto 1991, Miyoshi 1996, Sasao 1996, Kameya et al. 1998). The meaning of the acronym VERA also changed from "\underline{V}LBI for the \underline{E}arth \underline{R}otation study and \underline{A}strometry" to "\underline{V}LBI \underline{E}xploration of \underline{R}adio \underline{A}strometry." In order to perform differential VLBI with high sensitivity, the final VERA system is equipped with 2-beam antennas, which have two receiving systems on every dish (Kawaguchi et al., 2000). Using the 2-beam we can observe both target source and reference source simultaneously, thereby on-source times become a factor of $\sim 4 $ longer than those of switching differential VLBI observations. In addition, the simultaneous observations of pair sources are free from misconnections in visibility phase interpolation between time gaps caused by antenna nodding, giving better recovery in coherence of the data.\\
  A new idea often brings with it new disadvantages together with new advantages.
Admitting the advantage of simultaneous observations of pair sources with 2-beam,
 problems also have been recognized which degrade the performance for astrometric measurements (Miyoshi 2007).
 
 
 One of the serious problems of the 2-beam VERA lies in its method of calibrating the instrumental differences between the two beams. In case of switching differential VLBI, the amounts of calibrations in gain and delay are almost common to both sources. In case of differential VLBI with the 2-beam system, there exist instrumental delay and gain differences between them that must be calibrated.
 The higher the requirement for astrometric accuracy, 
the narrower become the margins for errors.
 For example, we can measure an annual parallax of 500 pc easily because the corresponding delay change is 4 cm for a 2000-km baseline while for measuring a 10 kpc distance by annual parallax, we must detect a 0.2 cm delay change for a 2000-km baseline.
Further,
 in order to establish high-precision astrometry of $10 \mu$arcseconds, precise calibrations become essential. If a $100 \mu$m of instrumental delay difference remains, it easily spoils the accuracy about $10\mu$arcseconds with a 2000-km baseline. In the same way, an error of 1 mm in a 1000-km baseline will cause a positional error of $10 \mu$arcseconds in $2^{\circ}$ separations measurement between a pair source.
 However, there is no method sufficiently accurate to calibrate the instrumental delay difference between the two beams within errors of 0.1 mm level. By the horn-on-dish method (Kawaguchi et al. 2000) we cannot measure all the instrumental delays formed as the result of the entire optics including the main reflector, because the radio wave from the emitters of the horn-on-dish method does not reflect on the main reflector, but goes directly to the receiving horns from the sub-reflector.
  There are at least two types of delays caused by the total optics. One is the instrumental delay difference occurring due to mispointing of sources. If antenna pointing has an offset larger than one-fifteenth of beam size, the phase in the shaped beam corresponds to a $50 \mu$m in length at 43GHz observations in case of $1^{\circ}$ separations.  Another is a convolution effect with the real shape of the reflector surface and an illumination pattern. As the receiving horn rotates on the dish to track the field of view, the phase of synthesized beam will change with the result of the convolution. Hence the horn-on-dish method is not sufficient to calibrate the whole instrumental delays (Miyoshi 2007).
 The supporting-antenna-method provides a unique possibility of measuring these real instrumental delay errors. By building an antenna close to the 2-beam antenna we can construct an almost zero baseline interferometer free from the atmospheric phase variation. By performing a switching observation of the paired-sources with the supporting-antenna, we can measure the instrumental delays between the supporting-antenna and each of the two beams. From the interferometric observations using the supporting antenna, we can indirectly measure the phase difference between the two beams (Miyoshi, 2004). At Mizusawa station, using the 10-m radio telescope we can perform the supporting-antenna-method though achieving this with the equipments in all stations is difficult because of high construction cost.

 From the viewpoint of radio interferometric techniques and calibration methods, it is technically interesting to check what will happen between the two beams and the degree to which we can calibrate the instrumental delay and gain errors in the 2-beam antenna. We also have an interest in the experiment reported by A2003 itself. The first report on the 2-beam antenna showed very high performance on correction of atmospheric phase variations even without the horn-on-dish method. A2003 showed the differential phases between the sources were zero on average with r.m.s. variations of $\pm30^{\circ}$. A2003 claimed that the resultant Allan standard deviations followed the theoretical curve, and the coherence function (vector averaging of complex visibilities) lasted almost 1.0 during 3600 seconds of integrations.
Further the usage of maser sources for testing interferometers is quite unique:
 A2003 used a water maser source as a target (OH~43.8-0.1) and also used a water maser as a reference source (W49N), which is usually not appropriate for testing performance of astrometric and interferometric instruments for the following reasons.

\begin{enumerate}
\item Maser sources often have complex structures. In such cases, the corresponding visibility phase includes components of the structure. If the structure consists of multiple points, which is frequent for interstellar maser structures, it will be difficult to remove the effect on phase variations.

\item The positional accuracies of maser spots are on the order of sub-arc to arcseconds in advance. If the tracking center has an offset of arcseconds in the correlator model, the raw visibility shows a large rate and the visibility phase will increase or decrease too rapidly to follow. In the worst case the coherence of the visibility data will be lost.

\item 
Though the observed maser source is strong, we cannot measure the delay accurately because the delay is measured from the phase slope with frequency. Maser sources have insufficient bandwidth for measuring the phase slope, and also the maser spots are often widely distributed on the sky, which means the geometrical delay has a different value in each frequency. 
\end{enumerate}

After all these difficulties, A2003 showed excellent results as if they had observed paired strong continuum point sources. We wondered what kind of good conditions occurred in the observations and decided to test the raw archival visibility data of A2003 directly in order to investigate the visibility in detail.\footnote{Signal received at a station is digitally recorded with precise time code. After observations the recorded signal is played back, corrected the geometrical delay difference calculated from earth rotation model, and cross-correlated with signal at other station in the correlator. We often call the cross correlation output visibility. The visibility is integrated for a short time, 1 or 2 seconds in correlator. In this paper we call such visibilities raw data because the amplitudes and phases of the visibilities are still not fully calibrated. The parameters used in correlation processing are rough, not enough for a fine synthesis imaging and a high precision astrometric measurement. Concerning the data processing of radio interferometry, see Thompson, Moran \& Swenson (2001) and other related text books. }

\section{Observation}
 A2003 noted two experimental observations, but mainly the second experiment was explained with figures  of visibility phases, Allan standard deviations, and coherence functions for 1-hour duration. We test the second epoch data set and show the details.
 The second observation was performed with one baseline between Mizusawa and Iriki, spanning about 1300 km. A pair of water maser sources, W49N and OH~43.8-0.1 was observed during about 6 hours, no strong continuum source was observed, which we usually add into observing schedules for calibrations of station clock parameters. The second observation spanned from 11:30 to 17:30 (UT) in July 2002, but the archival data exist until 15:35 (UT) for W49N correlations. We here test the data of the existing duration from 11:30 to 15:35 (UT).

\section{Analysis and Results}
 We show several aspects of the visibility data using tasks of the NRAO AIPS (Astronomical Image Processing System) package which is a world-wide common tool for calibration, data analysis, image display, plotting, and a variety of ancillary tasks on interferometric data.
 The AIPS is sufficiently reliable to check visibility data here, though not fully being testified for high precision astrometric measurements $\sim 10 \mu $ arc-seconds level.
We used the AIPS task POSSM that displays auto- and cross-power spectra of visibility data, got fringe search solutions with the task FRING, and tried to estimate positional offset from a multiple-point fringe-rate method with the task FRMAP.

  We then output the visibilities of the peak channels, and show the respective amplitude and phase variations, tried to correct the positional offset by the fringe phase mapping method, and finally calculate the Allan standard deviations of visibilities.

\subsection{Auto- and cross-power spectra with POSSM}
 Figure 1 shows spectra of each object, of both auto- and cross-power spectra. For the cross-power spectra of W49N, we corrected phase and rate using FRING solutions from the peak frequency channel 205 ch.
 Unfortunately we could not correct the delay due to the lack of continuum observations.
For the cross-power spectra of OH~43.8-0.1, we corrected the same amounts of rate and phase as those of W49N, and also applied a rough positional offset correction ($\Delta\alpha = -5", \Delta\delta = -5"$), obtained from FRMAP in AIPS.
The auto-power spectra of both stations are similar, suggesting equal sensitivities of the two antennas. The cross-power spectrum of W49N shows a very steep phase slope with frequency while that of OH~43.8-0.1 shows noisy features except around 135 ch.\\

\begin{figure}
\begin{center}
\FigureFile(8cm,8cm){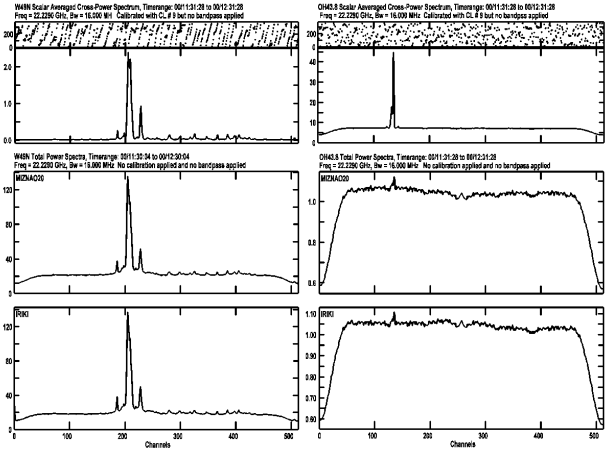}
\end{center}
\caption{Auto- and cross-power spectra of W49N and OH~43.8-0.1.
The left panels show those of W49N, and the right panels show those of OH~43.8-0.1.
The top panels show the cross-power spectra, 
the middle panels the auto-power spectra at Mizusawa station and
the bottoms those at Iriki station.
 The integration was done for the first hour, from 11:30 to 12:30.
Because the measured system temperatures are not certain today, the amplitude scales here are arbitrary, as shown as the values of raw visibilities.}
\end{figure}

\subsection{Fringe search parameters with FRING}
 Figure 2 shows the rate and phase solutions obtained by FRING from the peak frequency channel 205 ch of W49N. Because the observed source was not a continuum source we cannot perform a delay search. We only performed fringe rate and phase searches by FRING in AIPS.
  The obtained rate solution shows violent variation from -20 to + 40 mHz within 4 hours:
the fringe rate decreases before the southing (about 13:30 UT, 7200 seconds from the observing start time), and increases after the time with a different change rate: signs of the fringe rate were plus (+) from 11:30 to 12:45, minus (-) from 12:45 to 14:15, and plus (+) again after 14:15. In other words, sign changes of fringe rate, or changes in increase and decrease of fringe phase occurred twice within 4 hours. This is quite different from the usual variation due to positional offset of the observed source. In the case of positional offset the phase should show a sinusoidal variation with a period of one sidereal day: change of increase and decrease should occur every 12 sidereal hours.\\

\begin{figure}
\begin{center}
\FigureFile(8cm,8cm){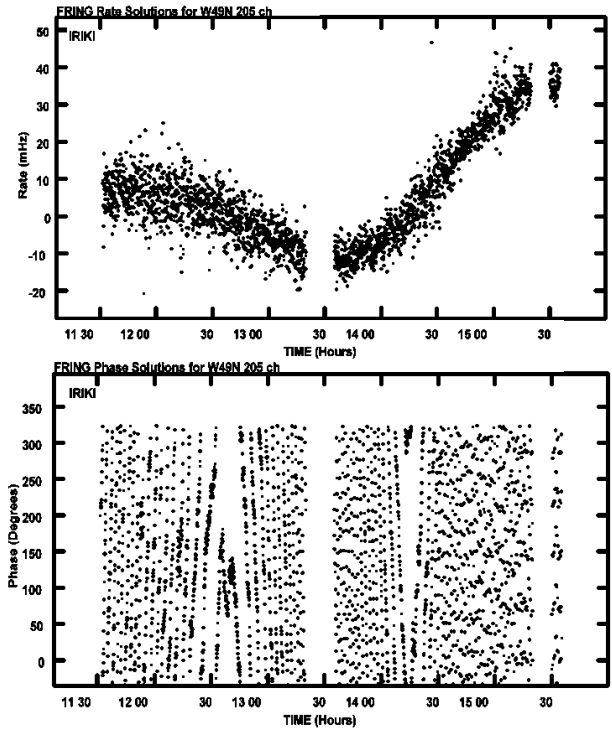}
\end{center}
\caption{Rate and phase solutions by FRING in AIPS from the peak frequency channel 205 ch of W49N.
}
\end{figure}

\subsection{An attempt to estimate a positional offset from multi-point fringe-rate method with FRMAP}

Using the task FRMAP in AIPS, we tried to get a positional offset between the real one and the assumed one applied in the correlator model.
The FRMAP in AIPS package is the task performing multiple-point fringe-rate method (Moran et al. 1968, Walker 1981, Thompson, Moran \& Swenson 2001).
 Residual fringe rates include information of positional offset of the source in the sky.
 In the early days of radio interferometers and VLBI, we frequently utilized fringe rates for obtaining observed source positions. The multi-point fringe-rate method is still useful today for measuring rough positions with accuracy reaching a few tens of milli-arcseconds in VLBI observations.
 Today, using radio interferometers like the Very Large Array (VLA), we can measure a maser position
 with an accuracy on the order of sub-arcseconds. However most maser sources are initially detected and observed by single dish with the positional accuracies of the order of arc-seconds.
 We sometimes perform VLBI observations of maser sources without more accurate positions, and
 find the positional offsets exist after correlation processing. The task FRMAP in AIPS is often used for finding positional offsets in such cases.

We tried to obtain the positional offset of OH~43.8-0.1 relative to W49N by FRMAP in AIPS utilizing the fringe rates of the 135 frequency channel in OH~43.8-0.1 after applying the FRING solutions obtained from W49N.
  Figure 3 shows that the estimation of the positional offset is not certain.
Not a single but several crossing points appear in FRMAP indicating existence of multiple maser spots with a rough positional offset of $ \Delta\alpha=5", \Delta\delta=5"$, though in usual case the positional offset should be more clearly indicated by exactly overlapping cross points. In this high signal-to-noise ratio case, the cross points are expected to concentrate within a sub-arcsecond or so.\\

\begin{figure}
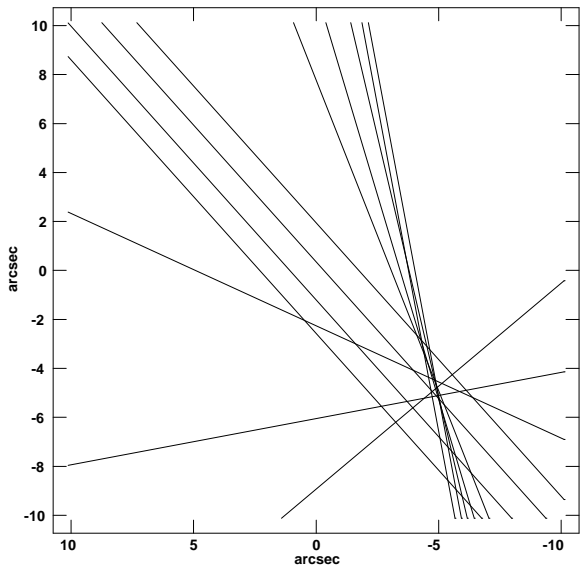

\begin{center}
\FigureFile(8cm,8cm){Figure3.eps}
\end{center}
\caption{Result of FRMAP concerning the frequency channel 135 ch of OH~43.8-0.1.}
\end{figure}
\subsection{Time variations of visibilities in the peak channels of W49N and OH~43.8-0.1} 
 For the second step, we output the visibility data set of the peak channels of both sources from AIPS' internal data format using the task UVPRT with a phase accuracy of $1^{\circ}$. Then we inspect the feature of visibility, showing the phase and amplitude variations with time. Finally, we try fringe phase mapping to the phase variations and attempt to remove the components of positional offset from phase after the $2\pi$n-connections of visibility phase.
 Figure 4 shows time variations of the visibilities of both sources: raw amplitudes, raw phases, $2\pi$n-connected phases, and residual phases after applying "apparent" positional corrections.
 

\subsubsection{Raw visibility time variations}
 The raw amplitude variations of the peak channels are shown in the top panels of Figure 4: 
the visibility of peak channel (205 ch) of W49N shows amplitude variation with ripples suggesting the existence of plural maser spots with different intensities (the left panel), while that (135 ch) of OH~43.8-0.1 shows very large variations reaching almost zero at minimum about 1000- to 2000- seconds intervals, suggesting that several maser spots with comparable flux densities exist.
 The raw phase variations of both sources are too rapid like a big downpour except around $t = 4000~and~10^4$ seconds in W49N and around $t = 4800$ seconds in OH~43.8-0.1. Also the changes between increase and decrease in phase variations occur twice in W49N, at least once in  OH~43.8-0.1 within 4 hours.

\subsubsection{Fringe phase mapping}
 As followed A2003, we try to estimate positional offsets of the individual sources and remove the effect from the phase variations. A2003 did not mention the details of how to subtract the effect. Here we use the classical and simple, fringe phase mapping method (Wade 1970, Thompson, Moran \& Swenson 2001). As well as the multi-point fringe-rate method, this method was frequently used to measure the source positions in VLBI.
 If the observed source has one component structure, the offset from the assumed position in correlator processing produces a sinusoidal phase change with a period of one sidereal day. Therefore we can estimate the positional offset from the sinusoidal phase curve. In principle this method cannot be applied to the case where the source has a complex structure because the phase does not show a simple sinusoidal variation owing to the effect of the source structure.
 As has already ben shown, the visibility data are not those of one component source. However, we tried this method in order to reproduce the results reported by A2003.

Before applying the fringe phase mapping method we must connect the phase beyond $\pm\pi$. We followed phase change by eye and connected the phase gap over $\pm\pi$.
 The panels third from the top in Figure 4 show the resultant $2\pi$n-connected phase in black lines. The $2\pi$n-connected phases of W49N show variations spanning about $3 \times 10^4 ~\degree$ (1 m in delay length) for 5 hours and the curve differs from a sinusoidal one with a period of one sidereal day. In connecting the phases of OH~43.8-0.1 we found several phase jumps of about $\pi$ which are due to the  beating of plural maser spots, where the $2\pi$n-connection was not perfect. The phase change of OH~43.8-0.1 shows large changes over $10^5 ~\degree$ ( 3.8 m in delay length) for 4 hours.

 We applied fringe phase mapping, namely fitting a sinusoidal curve with a period of one sidereal day to the variations of $2\pi$n-connected phases, an operation equivalent to the description in A2003 which says "we fitted the fringe phase (and also the fringe rate) based on the relation $\phi=U\Delta X + V\Delta Y$ ($\phi$ is the fringe phase, $U$ and $V$ are the projected baseline components in the UV plane, and $\Delta X$ and $\Delta Y$ are position offsets of tracking center)".
 
 We first performed fringe phase mapping on the whole time data set first: the red lines in the panels third from the top in Figure 4 are the resultant fit curves, and the panels fourth from the top show the residual phases after subtracting the amounts of the fitted curves, which show the large discrepancies between the fitting curves and the observed phase variations.
  Next we applied this method to a part of the data set i.e., the first one-hour duration with just the same span as A2003. Figure 5 shows the fitting result of the limited time. The residual phases of both sources and the differential phases during the first hour coincide with those reported in A2003. The variations of our residual phases after fitting are almost the same as A2003 reported. Also our differential phases, shown as a blue line in the bottom panel show almost zero on average for the first hour as reported by A2003. But the residual phases and the differential phases of the successive times show a big downpour feature. In other words, fitting curves to the first hour did not provide the real positional offset.\\

\begin{figure}
\begin{center}
\FigureFile(8cm,8cm){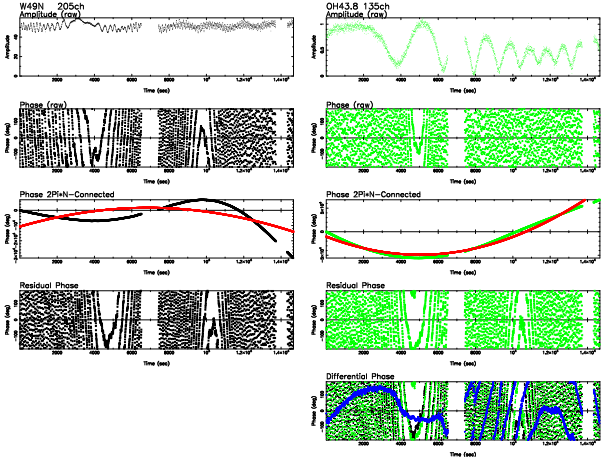}
\end{center}
\caption{Amplitude and phase variations with time: 
visibility of the 205 ch in W49N (left panels) and that of the 135 ch in OH~43.8-0.1 (right panels).
The top panels show the raw amplitude variations, the panels second from the top show the raw phase variations, the panels third from the top show the $2\pi$n-connected phase and the fitting curve of a sinusoidal curve with a period of one sidereal day (red lines). 
The panels fourth from the top show the residual phases after subtracting fitting curves. The bottom right panel shows the differential phase (blue) and the residual phases (black for W49N, green for OH~43.8-0.1).
}
\end{figure}

\begin{figure}
\begin{center}
\FigureFile(8cm,8cm){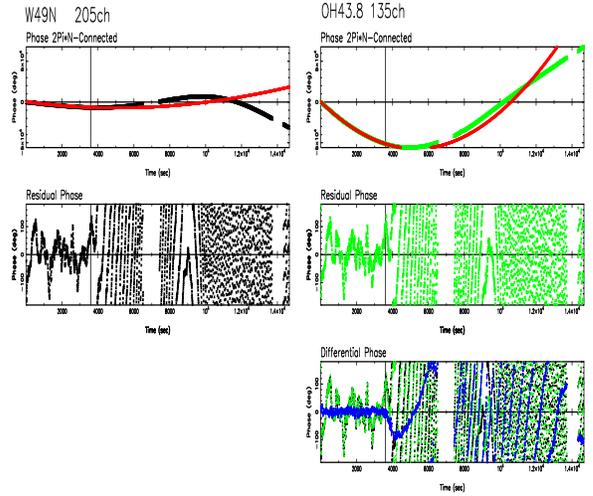}
\end{center}
\caption{Phase variations and the results of the first hour fitting.
Visibility of the 205 ch in W49N (left panels) and that of the 135 ch in OH~43.8-0.1 (right panels). The top panels show the $2\pi$n-connected phase and the fitting curve of a sinusoidal curve with a period of one sidereal day (red). The panels second from the top show the residual phases after subtracting fitting curves. The bottom right panel shows the differential phase between the residual phases (blue) and the residual phases (black for W49N, green for OH~43.8-0.1).
}
\end{figure}

\subsubsection{Allan Standard Deviations} 

 Here we show several Allan standard deviations (ASDs) of the residual phases after subtracting different fitted functions in order to compare the effects of fitting functions.
 In Figure 6, the black curve (Raw) shows the Allan standard deviations of the pure differences between raw visibility phases that are the correct ASDs we should calculate for the investigation, which decreases in the region of $\tau < 40$ seconds showing "white noise features" but increases in the region of $\tau > 40$ seconds.
 The black line (Line) shows the ASDs of the residual phases after subtracting values of a fitted linear function to the first one-hour visibilities. The blue curve (Quad) shows the ASD after subtracting those of the fitted quadratic function. The red curve (Sin24) is the ASDs after subtracting those of the fitted sinusoidal function with a period of one sidereal day that A2003 calculated and showed in their Figure 2 presumably with a misunderstanding what ADSs are. The curves ASDs of the Quad and Sin24 deviate from those of Raw and Line when the $\tau$ is larger than 40 seconds.\\
\begin{figure}
\begin{center}
\FigureFile(8cm,8cm){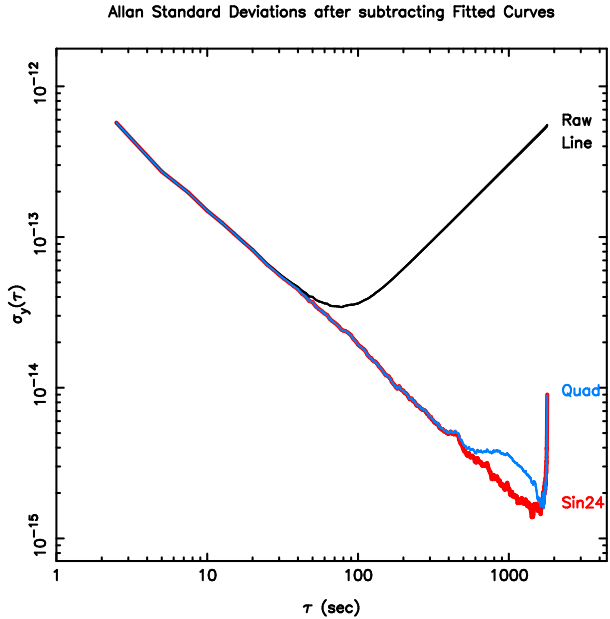}
\end{center}
\caption{Allan standard deviations (ASDs).
A black curve denoted by Raw shows the ASDs of the differential phases of the two sources, which are exactly the same as the ASDs of the residual phase after subtracting the linear function fitted to the raw data for the first 1 hour. The red curve denoted by Sin24 shows the ASDs of the residual phase after subtracting a sinusoidal function with a sidereal day period to the raw data for the first 1 hour.
The blue curve denoted by Quad shows the ASDs of the residual phase after subtracting a quadratic function fitted to the raw data for the first 1 hour.
}
\end{figure}

\section{Discussion}

 A2003 claimed to confirm performance of the VERA 2-beam system for corrections of atmospheric phase variations and the capability of maintaining good coherence allowing a longer time integration. However, the concerned archive data show features different from the arguments and conclusions of A2003.

\begin{enumerate}
\item
A2003 showed only the first hour data and described the performance and capability
of the instruments. We found the existence of successive 5 hour data; the observations were 6 hours in all. However, there was no mention of the successive 5 hours in A2003 though the data show high signal-to-noise ratios.

\item
We found the "apparent" positional offset correction only valid for the first hour
 and not valid for other successive times.
There occur very large and rapid phase variations during the successive 3 hours in spite of applying the "apparent" positional offset correction.
Therefore the value of "apparent" positional offset correction is not the real positional offset of the observed sources.

\item
By investigating the details of all phase variations, we found a very rapid phase change at the southing time. Also the phase variations differ from any sinusoidal variations with one sidereal day period, which should be due to incorrect functioning of the system at the time the experiment was performed. In fact, errors were found in the correlator model after publication of A2003 (Shibata et al. 2004). For more than one year after the publication of A2003, unknown errors in the correlator models hindered us from getting synthesis maps by phase referencing while we got individual images from usual hybrid mapping method. Later the reason was found to be errors in calculations of the tracking delay and delay rate parameters in the Mitaka FX correlator: the atmospheric delay term becomes infinite for the zenith observations because of mistreatment of an elevation angle as a zenith angle together with the inversed sign of the term. The pair of W49N and OH~43.8-0.1 almost reached zenith at their southing time. The behavior of the large phase change at the southing found in our reanalysis can be explained by the error in atmospheric delay term in the correlator.

\item
 The amplitude variations of the visibility of both sources show that the source structures are not a single compact one. Also the fringe rate mapping shows the possibility that the maser distribution is in arcseconds order. This means that the waves of the observed radio emissions are not in plane-parallel, not appropriate for testing performances of VLBI.
 As noted in the first section, the maser sources are not proper for checking and testing the performances of VLBI or for investigating the atmospheric phase fluctuations because of their complicated structures. The corresponding fringe phases include information of the structure, therefore it is difficult to extract the geometric and instrumental delay phases from the obtained phases.
Also because the maser emission is in a narrow bandwidth, we cannot measure the absolute delays by using group delay (= phase slope with respect to frequency).

\item
We finally discuss the treatment of Allan standard deviations.
In general, the Allan standard deviations of different time series data are different from each other.
So the subtraction of value of fitted function to the data before calculating Allan standard deviations will replace the value and nature of Allan standard deviations.
Hence the value and nature of the Allan standard deviation shown in A2003 are not necessarily the same as those of the raw visibility data and pure phase differences between the two sources. As shown in Figure 6, the Allan standard deviations derived from the A2003 manner are different from those of the raw differential phases as long as the $\tau$ is longer than 40 seconds.
 Allan standard deviations belong to operations of the second finite differences with respect to time.
No change will occur in the Allan standard deviations if we subtract values of functions whose second time derivatives are always zero. We can safely subtract values of a linear function from original time series data before calculating Allan standard deviations. Therefore the Raw and LINE in Figure 6 show the same ASDs, while A2003 performed subtraction by a fitted sinusoidal function. Because the second difference of a sinusoidal function is not always zero, the subtraction from the raw data causes a change in the nature of Allan standard deviation from the original one. In such a case we should evaluate the effect of fitting functions to the ASDs qualitatively. As shown Figure 6, the ASDs of Quad and Sin24 are quite different from those of raw data, presumably the "high pass filtering" in A2003 is only valid where $\tau$ is less than 40 seconds. 
\end{enumerate}

 In A2003, after applying the "apparent" positional offset correction the differential phases showed zero on average with r.m.s. of about $\pm 30 \degree$ for the first hour: also the coherence kept almost 1 during the first hour and the Allan standard deviations showed white noise features for longer times than 40 seconds.
 A2003 claimed that the features indicate the successful correction against atmospheric phase variations  by the 2 beam simultaneous differential VLBI and that the possibility for longer integration was demonstrated by the data. Admitting the effect of 2-beam simultaneous observations on coherence for shorter time scales of less than 40 seconds, we found the recovered coherence is not from the 2-beam observations, but mainly from the effect of curve fitting to only the first hour.


\section{Conclusion}
We found that the data of the first 2-beam VERA observations reported by A2003
do not show conclusive results about the performance of the interferometer for atmospheric phase corrections. Rather the data indicate that the instrument including the correlator contained some errors at that time. Thereby the data lost the precise positional information of the observed sources. In fact, Shibata et al. (2004) found the errors in the correlator model after the publication of A2003. If they checked the whole data set obtained by A2003 observations, they would easily have found the large and rapid phase variations caused by errors in correlator models.
 
 Further, the Allan standard deviations of "the differential phase" shown in A2003 are not those of the observed differential phases, because their fitting action to the observed visibility is neither a positional correction nor a high-pass filtering of shorter-term phase variation.  Their fitting action in A2003 altered the nature of Allan standard deviations in longer-term than 40 seconds.
 
  The results shown in A2003 are not proper for discussing the performance of phase correction by the 2-beam method. The real and recent performance of the instrument should be evaluated with other data sets using not maser sources but compact continuum sources. Investigations of atmospheric phase variations like those shown in A2003 are interesting and important for phase correction techniques in radio interferometers. A new performance test of the instrument is also expected from this point of view.



\begin{thebibliography}{}
\bibitem[Fujishita, \& Hara 1988]{}
Fujishita, M., and Hara, T., 
The Impact of VLBI on Astrophysics and Geophysics;
Proceedings of the 129th IAU Symposium, Cambridge, MA, May 10-15, 1987.
Edited by Mark Jonathan Reid and James M. Moran.
Symposium sponsored by IAU, URSI, NASA, et al. Dordrecht, Kluwer Academic Publishers, 1988., p.483

\bibitem[Hara 1986]{}
Hara, T., Proceedings of the Symposium on Application of Space Techniques to Astronomy and Geophysics,
, 1986, p. 160 - 163

\bibitem[Hara et al. 1988]{}
Hara, T., Okamoto, I., and Sasao, T. 
Vistas in Astronomy (ISSN 0083-6656), vol. 31, 1988, p. 647-652.

\bibitem[Honma et al. 2003]{}
Honma, M., et al. 2003, PASJ, 55, L57-L60

\bibitem[Kameya et al. 1998]{}
Kameya, O., Sasao, T., and Miyoshi, M. 
The central regions of the Galaxy and galaxies, Proceedings of the 184th symposium of the International Astronomical Union,
held in Kyoto, Japan, August 18-22, 1997. Edited by Yoshiaki Sofue. Publisher: Dordrecht: Kluwer, 1998. ISBN: 079235060X, p.327

\bibitem[Kawaguchi, Sasao \& Manabe 2000]{}
Kawaguchi N., Sasao T., Manabe S. 2000, in Proc. SPIE Vol.4015 Radio Telescope, ed H. R. Buthcer, p544 - p551







\bibitem[Miyoshi 1996]{}
Miyoshi, M. 1996, in proceedings of the third East-Asian meeting on astronomy, July 17-21, 1995, Tokyo, Japan. Edited by Norio Kaifu. Tokyo, Japan: National Astronomical Observatory, 1996, p.481

\bibitem[Miyoshi 2004]{}
Miyoshi, M. 2004, Japanese Patent pending
\bibitem[Miyoshi 2007]{}
Miyoshi, M. 2007, Report of the National Astronomical Observatory of Japan, 10, 1-18

\bibitem[Moran et al. 1967]{}
Moran, J. M., Burke, B. F., Barrett, A. H., Rogers, A. E. E., Ball, J. A., Carter, J. C., Cudaback, D. D.,\& Walker, R. C., 1981, ApJ, 86, 1323-1331

\bibitem[Sasao, \& Morimoto 1991]{}
Sasao T. \& Morimoto M., Antenna cluster - Antenna cluster VLBI for Geodesy and Astrometry, NOAA Technical Report NOS 137 NGS 49 Proceedings, Chapman Conference on Geodetic VLBI: Monitoring Global Change, 48-62, 1991

\bibitem[Sasao 1996]{}
Sasao T. 1996, in proceedings of 4th APT Workshop, ed. E. A. King, p94 - p 104

\bibitem[Shibata et al. 2004]{}
 Shibata, K. M., Tamura, Y., Bushimata, T., Jike, T., and Kobayashi, H. 2004,
The status of the Mitaka correlation center,  
The proceedings of the Japan VLBI consortium symposium (Dec. 2003), p175 

\bibitem[Thompson, Moran \& Swenson 2001]{}
Thompson A. R., Moran J. M., Swenson G. W. Jr. 2001, Interferometry and Synthesis
 in Radio Astronomy 2nd edition (Wiley-Interscience, New York)

\bibitem[Wade 1970]{}
Wade, C. M., 1970, ApJ, 162, 381-390

\bibitem[Walker 1981]{}
Walker, R. C., 1981, ApJ, 86, 1323-1331
\end{thebibliography}
\end{document}